\documentclass[pre,twocolumn,unsortedaddress,floatfix,amssymb,showpacs]{revtex4-1}
\usepackage{times}
\usepackage{graphicx}
\usepackage{amsmath}
\usepackage{color}
\usepackage{natbib}
\bibpunct{[}{]}{,}{n}{}{;}

\begin{document}

\title{The dynamics of polymer ejection from capsid}

\author{R. P. Linna}
\email{Author to whom correspondence should be addressed: riku.linna@aalto.fi}
\author{J. E. Moisio}
\author{P. M. Suhonen}
\author{K. Kaski}
\affiliation{
  Department of Biomedical Engineering and Computational Science,
  Aalto University, P.O. Box 12200, FI-00076 Aalto, Finland
  }


\begin{abstract}
Polymer ejection from a capsid through a nanoscale pore is an important biological process with relevance to modern biotechnology. Here, we study generic capsid ejection using Langevin dynamics. We show that even when the ejection takes place within the drift-dominated region there is a very high probability for the ejection process not to be completed. Introducing a small aligning force at the pore entrance enhances ejection dramatically. Such a pore asymmetry is a candidate for a mechanism by which a viral ejection is completed. By detailed high-resolution simulations we show that such capsid ejection is an out-of-equilibrium process that shares many common features with the much studied driven polymer translocation through a pore in a wall or a membrane. We find that the escape times scale with polymer length, $\tau \sim N^\alpha$. We show that for the pore without the asymmetry the previous predictions corroborated by Monte Carlo simulations do not hold. For the pore with the asymmetry the scaling exponent varies with the initial monomer density (monomers per capsid volume) $\rho$ inside the capsid. For very low densities $\rho \le 0.002$ the polymer is only weakly confined by the capsid, and we measure $\alpha = 1.33$, which is close to $\alpha = 1.4$ obtained for polymer translocation. At intermediate densities the scaling exponents $\alpha = 1.25$ and $1.21$ for $\rho = 0.01$ and $0.02$, respectively. These scalings are in accord with a crude derivation for the lower limit $\alpha = 1.2$. For the asymmetrical pore precise scaling breaks down, when the density exceeds the value for complete confinement by the capsid, $\rho \gtrapprox 0.25$. The high-resolution data show that the capsid ejection for both pores, analogously to polymer translocation, can be characterized as a multiplicative stochastic process that is dominated by small-scale transitions. 
\end{abstract}

\maketitle

\section{Introduction}
Polymer ejection from inside a capsid through a small pore is an important biological process~\cite{alberts}. Viral packaging in and ejection from a bacteriophage are extensively studied processes~\cite{muthukumar1,muthukumar2,smith,grayson,ghosal,cacciuto2,sakaue3}. It is well established that {\it e.g.}\ a double-stranded (ds) DNA assumes a spooled conformation inside a capsid~\cite{ghosal,cerritelli,kindt,purohit}. In contrast, a fully flexible polymer assumes a random conformation. Examples of such polymers are RNA, single-stranded (ss) DNA, and proteins. They are flexible compared with dsDNA and can often be modeled as freely jointed chains (FJC) having no bending rigidity. The ejection dynamics of a polymer starting from an initially spooled and from a random conformation inevitably differ, for example due to the constant rotary motion required for the spooled polymer to eject. 

Capsid ejection has sometimes been studied as a special case of confinement-driven ejection. Experimentally, the roles of the confinement-entropic force and entropic elasticity have been characterized~\cite{turner,krasilnikov}. Polymer ejection from a long pore or a tube has been extensively studied, see {\it e.g.}~\cite{prinsen}, where the long-standing problem of how the genome delivery process is completed after the initially higher pressure drops inside the capsid during ejection was addressed. Some exact results were derived and obtained via Monte Carlo simulation in~\cite{milchev}. Although general characteristics of capsid ejection can to some extent be addressed by investigating ejection from {\it e.g.} a tube, it is evident that a detailed understanding requires the exact geometry of the confinement. The importance of the shape of the confinement was shown in a computational study comparing  ejection from ellipsoid-shaped and spherical capsids~\cite{ali}, which is in agreement with the notion that the force driving the polymer out of confinement exhibits a nontrivial dependence on geometry~\cite{grosberg,cacciuto}. Obviously, the solvent quality affects the ejection process. This was investigated in~\cite{ali2}.

In order to understand the biologically relevant process of a polymer ejecting from a a capsid it is important to have a correct view of the underlying generic process. Here the polymer starts from a random confined conformation. Ejection can be viewed as a special case of polymer translocation. For clarity, in what follows, we will call {\it translocation} the process of a polymer moving from one half space to another through a nanometer-scale pore in the wall or membrane separating them and refer to the process of polymer exiting a capsid through a nanometer-scale pore as {\it capsid ejection}.

The relation between capsid ejection and polymer translocation is of particular importance. In the early important work by Muthukumar on the ejection of a polymer from a capsid starting from a random conformation~\cite{muthukumar1} and the following reinvestigation by Cacciuto and Luijten~\cite{cacciuto2} the Monte Carlo method was used. The formalism invoked to explain the findings assumes that the polymer, initially in spherical confinement, moves slowly enough to remain in equilibrium. According to blob scaling result~\cite{degennes}, in~\cite{muthukumar1} the initial free energy of the confined polymer was assumed to scale as $F \sim N/R^{1/\nu}$, where $N$ is the number of monomers, $R$ the radius of the confining sphere, and $\nu$ the Flory exponent. In~\cite{cacciuto2} it was noted that the correct initial scaling for the spherical confinement is $F \sim N \phi^{1/(3\nu-1)}$, where $\phi$ is the monomer volume fraction. Using this scaling relation and the lower bound for the translocation time, derived in~\cite{kantor}, the authors obtained the scaling relation 
\begin{equation}
\tau \sim N^{1+\nu}\phi^{1/(1-3\nu)}
\label{ref_scaling}
\end{equation}
for the ejection time.

Sakaue and Yoshinaga presented an analytical formulation for the capsid ejection process~\cite{sakaue3}. They noted that the origin of the scaling in Eq.~(\ref{ref_scaling}) is unclear. Most notably, they criticized the proposition that a polymer is driven by the strong and constant free energy of the confinement. In their analysis the driving force diminishes as the monomer density inside the capsid drops. Accordingly, the ejection is first dominated by the osmotic driving force and later by the diffusion of the polymer. The resultant osmotic force is exerted only on the monomer residing at the pore, which induces a response that depends on the monomer's distance from the pore. No assumption of the polymer remaining in equilibrium is made in this description. Related to this, it has recently been shown that the experimentally measured variation of the mobility of the ejecting DNA with its remaining length in the capsid~\cite{grayson} can be explained by considering the free energy of the spooled conformation and the friction between sliding DNA strands and the capsid wall~\cite{ghosal}. Similarly to the work in~\cite{sakaue3}, this approach does not invoke any assumptions about the polymer remaining in or close to equilibrium or being ejected rapidly straight from the initial conformation.

We have previously shown that driven polymer translocation is a strongly out-of-equilibrium process~\cite{linna1,linna2}, where the polymer is continuously lifted further out of equilibrium during translocation so that on the {\it cis} side there is an increasing region where the polymer is under tension and the monomers are in motion. This idea was earlier used by Sakaue in his analytical calculation~\cite{sakaue}. The analytical treatment has been adopted and expanded in~\cite{dubbeldam}. Sakaue's concept was given further confirmation by a generalized computational model~\cite{ikonen}. The model was improved in~\cite{sakaue2}. We have previously reported that for a pore of finite friction the scaling $\tau \sim N^{1+\nu-\chi}$ is obtained and that for a pore of zero friction the scaling is $\tau \sim N^{1+\nu}$ for large enough pore force~\cite{linna1}. Rowghanian and Grosberg have derived a comprehensive and quite conclusive theory for polymer translocation in the asymptotic limit of very long polymers~\cite{grosberg2}. In this asymptotic limit the translocation was confirmed to scale as $\tau \sim N^{1+\nu}$. In all computational work the polymers are inevitably well below the length required to obtain asymptotic scaling. A finite-size scaling presented by Ikonen {\it et al} shows the close connection of zero pore friction and the asymptotic limit~\cite{ikonen2}.

Apparently, in the Monte Carlo simulations the ejection rate was sufficiently high for obtaining statistically meaningful results even at fairly low initial monomer densities inside the capsid~\cite{muthukumar1,cacciuto2}. This is somewhat unexpected, since there is the long-standing problem of how the ejection is completed~\cite{prinsen,lemay}. One explanation for the problem {\it in vivo} is offered by particles binding to the translocating or ejecting polymer. These chaperones inhibit the backsliding of the polymer~\cite{simon,sung,zandi,abdolvahad}. The binding particles acting only, or at least more actively, on the {\it trans} side in effect introduce an asymmetry across the pore, which we will model in a simple way. Recently, osmotic and hydrostatic pressure induced water flow was argued to be the mechanism through which the genome ejection is completed~\cite{lemay}. For a recent review on phage genome ejection, see~\cite{molineux}. 

The strong effect the pore friction has on the capsid ejection~\cite{grayson,ghosal} strongly suggests that the polymer is driven increasingly out of equilibrium during ejection just as in the case of driven polymer translocation. The importance of the pore friction is generally agreed upon in polymer translocation.  In accordance with the strong non-equilibrium nature of the translocation process it has been found that the dynamics used may have influence on the outcome in polymer translocation. In particular, the results of Monte Carlo simulations were found to deviate from those of Langevin dynamics simulations. This deviation increased with increasing pore force~\cite{linna2}. In the case of different viscosity on either side of the separating wall it was shown that using strict Brownian dynamics gives the translocating polymer a bias that is opposite to the bias given by Langevin dynamics~\cite{dehaan}. In capsid ejection the effective viscosity inside the capsid is higher than outside. These two findings and the notion of the potential nonequilibrium nature of the process together with the question of how the capsid ejection process is completed call for a reinvestigation of the capsid ejection process using a simulation method that captures the true dynamics of the process.

The paper can be outlined as follows. First, in Sect.~\ref{model} we describe our computational model. In Sect.~\ref{res} we present the results for the capsid ejection and compare them to those for polymer translocation. In Sect.~\ref{concl} we summarize our findings and draw conclusions.

\section{The computational model}
\label{model}

In order to make a close comparison between capsid ejection and polymer translocation we use dynamics and polymer model identical to those we used in Ref.~\cite{linna3}.

\subsection{The polymer model}

The standard bead-spring polymer model is used. Here, 
adjacent monomers are connected with anharmonic springs, described by the  finitely
extensible nonlinear elastic (FENE) potential 
\begin{equation}
\label{fene}
U_F = - \frac{K}{2} R_0^2 \ln \big ( 1- \frac{r^2}{R_0^2} \big ),
\end{equation}
where $r$ is the length of an effective bond and $R_0 = 1.5 \sigma$ is the maximum bond
length. The shifted Lennard-Jones (LJ) potential
\begin{equation}
\label{lj}
U_{LJ} = 4 \epsilon \left[ \left(\frac{\sigma}{r}\right)^{12} -
\left(\frac{\sigma}{r}\right)^{6} + \frac{1}{4} \right], \: r \leq 2^{1/6},
\end{equation}
is applied between all beads with distance $r$ apart. The parameter values were chosen as $\epsilon = 1.0$, $\sigma = 1.0$ and $K = 30 / \sigma^2$. The length scale can be related to physical length scale for instance by the relation $b = 2 \lambda_p$, where $b = 1$ is the bond length and $\lambda_p$ is the persistence length, {\it e.g.} $40$ \AA \ for ssDNA~\cite{tinland}. Unless otherwise noted, all lengths are given in units of $b = 1$.

\subsection{The dynamics}
\label{dn}

The dynamics of the polymer translocation was simulated by using Ermak's implementation of Langevin dynamics~\cite{ermak}. Accordingly, the time derivative of the momentum of the polymer bead $i$ is given by
\begin{equation}
\label{brownian}
\dot{\mathbf{p}}_i(t) = - \xi \mathbf{p}_i(t) + \eta_i(t),
\end{equation}
where $\xi$, $\mathbf{p}_i(t)$, and $\eta_i(t)$ are the 
friction constant, momentum, and random force of the bead $i$, respectively. The time 
integration was performed using the velocity Verlet algorithm~\cite{vangunsteren}. We set the Boltzmann constant $k_B = k = 1$. The temperature used in the simulations is $T^* = kT/\epsilon = 1$. The forces in reduced units are $\vec{f}^* = \vec{f}\sigma/\epsilon = \vec{f}$. $\xi = 0.5$, and $\eta_i(t)$ is related to $\xi$ by the fluctuation-dissipation theorem. The mass of a polymer bead is $m = 16$. The time step used in the numerical integration of the equations of motion is $\delta t^* = (\epsilon/{m \sigma^2})^{1/2} \delta t = 0.001$~\cite{allen}. Times are given in units of $\delta t^*$.

\subsection{The capsid and the pore}
\label{cp}

The spherical capsid has walls of thickness $3 b$, where $b = 1$ is the Kuhn length for the model polymer. Momentum reversal in the direction perpendicular to the capsid wall on the polymer beads hitting the inner or outer surface prevents them from entering the wall. In the directions tangential to these surfaces no-slip boundary conditions are applied. The pore is modeled as a cylindrical potential whose center axis is perpendicular to the wall and extends through it. Hence, the pore length is also $3 b$. The pore diameter is $1.2 \sigma$. Inside the pore, the  cylindrically symmetric damped linear force pulls the beads toward the pore axis 
\begin{equation}
f_h = -k r_p - c v_p, 
\end{equation}
where $k = 1000$, $c = 1$, $r_p$ is the distance of a polymer bead from the pore axis, and $v_p$ is its velocity component perpendicular to the axis.

The cylindrical pore is a versatile pore model that enables one to precisely determine the effective pore friction. The combination of no-slip boundary conditions on the capsid wall and the aligning force inside the pore can create a situation, where the bond between the bead inside the pore and its adjacent bead outside it crosses the corner where the capsid and the walls meet. Hence, the bead outside the pore can get stuck due to the no-slip condition and the bond may break. Such local effects are by no means unphysical. In the present study we are interested in generic (scale-invariant) features of the ejection process and hence have implemented an aligning local force exerted on the first bead outside the pore in the direction perpendicular to the pore axis. The magnitude of this small force is determined by the fraction $\mu$ of the crossing bond inside the pore, $f = - \mu k r_p$. Hence, the force magnitude is equal to what would be exerted on the polymer chain if the pore potential interacted with a continuous charge distibution along the chain. This is close to the actual situation. In the  DNA the charges reside along the chain at very small intervals compared to the pore dimensions. This aligning force is not necessary at the pore exit on the outer surface of the capsid. As will be seen, applying this force only at the entrance and not at the exit of the pore has a profound effect on the ejection process.

\subsection{Initial conformations}

\begin{figure}[t]
\centerline{
\includegraphics[angle=0, width=0.11\textwidth]{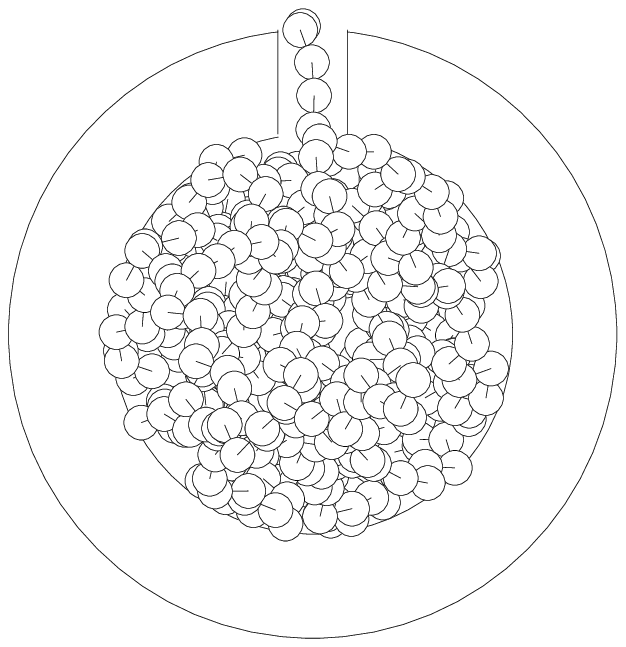}
\includegraphics[angle=0, width=0.15\textwidth]{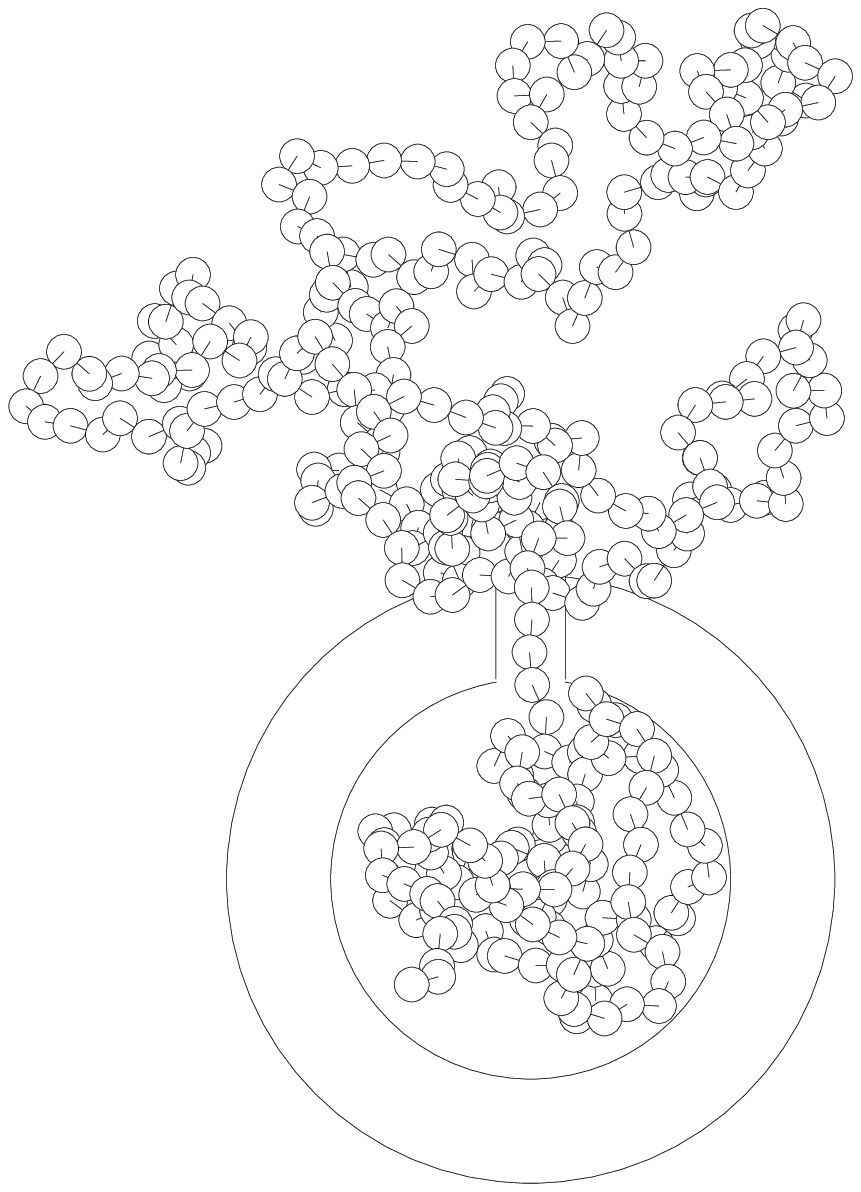}
\includegraphics[angle=0, width=0.15\textwidth]{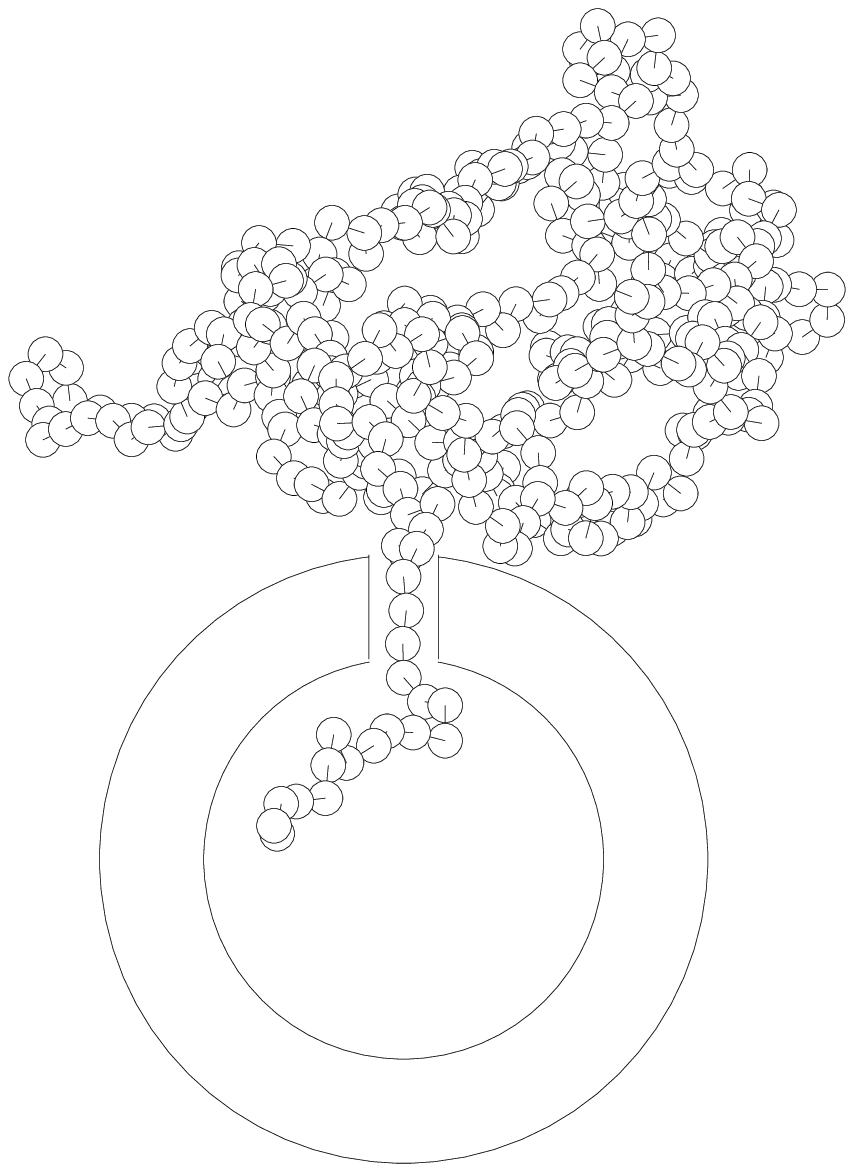}
}
\caption{Snapshots of a polymer of length $N=404$ escaping through the asymmetrical pore from inside a capsid of radius $R = 5.759$ ($\rho = 0.5$).}
\label{fig1}
\end{figure}

The polymers were either directly generated in the capsid or, in the case of dense packing, first generated outside the capsid with one end inside the pore and then packed inside by applying a constant force on the polymer segment in the pore. One polymer bead (monomer) was initially placed outside the capsid and three beads were placed in the pore. Hence, when investigating the ejection of a polymer {\it e.g.}\ of length $400$ we actually have $404$ beads of which $400$ are initially inside the capsid. Polymers were let relax to obtain the initial equilibrium conformation before allowing them to start escaping through the pore. Figure~\ref{fig1} shows the initial conformation of $400$ beads inside a capsid of radius $5.759$, which corresponds to the initial monomer density, {\it i.e.}, monomers per capsid volume, $\rho = 0.5$ inside the capsid.

\section{Results}
\label{res}

\subsection{Symmetrical pore}
\label{sp}

As described in Sec.~\ref{cp}, the cylindrical pore model requires a local alignment force at its entrance in order to avoid jamming of the polymer there. This force is a small fraction of the force exerted on a bead inside the pore. It is applied in the direction perpendicular to the pore axis and hence does not directly affect the momentum in the direction of ejection. The alignment is not necessary at the exit of the pore, nor is it needed in polymer translocation, where monomer densities at the pore entrance are lower. However, to preserve the symmetry of the pore model we first apply the alignment at both ends of the pore.

For monomer densities $\rho \le 0.2$, very few initial polymer conformations lead to ejection. The ejection probability for $\rho = 0.2$ is approximately $0.6$ \%. Moreover, the ejection process is very slow. These two factors prohibit gathering sufficient statistics from simulations for these densities. For $\rho = 0.1$ the capsid radius $R$ is of the order of the radius of gyration $R_g$ of the free polymer, which according to~\cite{cacciuto2} defines the border between the entropic barrier dominated diffusion- and the drift-dominated regimes. The ejection probablity increases abruptly at $\rho = 0.25$, where it is $50$\%-$60$\%. For this density the ejection rate is barely high enough for simulations for polymers of length $N = 200$ to complete in $21$ days, which is the maximum runtime limit in the supercomputers we have access to. For $\rho = 0.5$ the ejection probability ranges from $80$\% to $95$\% depending on $N$. For this density the ejection rate is barely sufficient for simulations for polymers of $N = 400$ to complete within $21$ days. Figure~\ref{fig2}(a) shows the ejection time $\tau$ versus the polymer length $N$ for $\rho = 0.25$ and $0.5$. $\tau$ scales with $N$, $\tau \sim N^{\alpha}$, where $\alpha \approx 2.06$ and $\approx 1.67$ for $\rho =0.25$ and $0.5$, respectively. These differ from the expenent $\alpha = 1.6$ argued by Eq.~(\ref{ref_scaling}).

As changing $\rho$ changes the way $\tau$ scales with $N$, so is changing $N$ expected to change how $\tau$ scales with $\rho$. The ejection times for only two densities, $\rho = 0.25$ and $0.5$, do not allow determining the relation between $\tau$ and $\rho$. However, they suffice to show that $\tau$ does not scale with $\rho$ according to Eq.~(\ref{ref_scaling}). For the measured $\nu \approx 0.6$, $\tau$ should scale as $\tau\sim \rho^\psi$, where $\psi = 1/(1-3 \nu) = -1.25$. Our measurements would give values $\psi \approx -0.93$, $-1.64$, and $-1.93$, for $N = 25$, $50$, and $100$, respectively.

The scaling $\tau \sim N^{\alpha}$ with $\alpha \approx 2.06$ for $\rho = 0.25$ is close to the scaling for unforced translocation, $\alpha \approx 2.2$, so one might think that the ejection approaches the limit where confinement has only a weak influence on the ejection dynamics. However, the capsid does confine polymers at $\rho = 0.25$, as can be seen in Fig.~\ref{fig2}(b), where the measured $R_g$ scales spherically with $N$, which indicates complete confinement of the polymer.

As we expect the ejection to be influenced by nonequilibrium effects, it is reasonable that the scaling argued by Eq.~(\ref{ref_scaling}) does not hold. It is well established that due to the strong out-of-equilibrium character of driven polymer translocation finite-size effects are present for the polymer lengths used in simulations. For similar reasons, the ejection dynamics is also expected to show finite-size effects for the polymer lengths used in simulations, and the scaling of $\tau$ with $N$ and $\rho$ cannot be expected to be obtained independently.

In~\cite{linna1} we used the evolution of the radius of gyration on the {\it trans} side $R_g^{tr}$ as a function of reaction coordinate $s$, {\it i.e.}, the number of translocated monomers, to show that the polymer on the {\it trans} side also is driven out of equilibrium during translocation. For the ejection $\rho = 0.5$ the {\it trans} side does not appear to be as strongly out of equilibrium, since $R_g^{tr}(s)$ follows fairly closely to the $R_g^{eq}(N)$ of equilibrated polymers of $N=s$. Accordingly, the measured $R_g^{tr}(s) \sim s^{0.6}$. In contrast, for the {\it asymmetrical pore}, discussed in Sect.~\ref{ap}, $R_g^{tr}(s) < R_g^{eq}(N=s)$, and the {\it trans} side is driven out of equilibrium during ejection in this case.

\subsection{Asymmetrical pore}
\label{ap}

Introducing a small asymmetry in the pore affects the ejection dynamics profoundly. As explained in Sect.~\ref{cp}, a small force is exerted on a single bead that is inside the capsid and connected to a bead inside the pore. The force is directed towards the center axis of the pore and its magnitude corresponds to the force that would be exerted on the part of the bond that is inside the pore. The resulting force is thus physically justified. However, implementing this only on the inside and not on the outside of the capsid introduces a small asymmetry: The beads just outside the capsid have a slightly smaller probability to slide back into the pore compared to the case where the alignment is done outside the capsid also. This effect is analogous to binding proteins that attach to translocated polymers preventing them from sliding back. In the asymmetrical pore the back sliding is made less probable, whereas the binding particles prevent the backsliding completely. The backsliding probability in the latter case is varied by the rates at which particles are attached to and unattached from the polymer~\cite{zandi}.

The most striking effect the pore asymmetry has is that polymers starting from very low monomer densities get ejected with high probability. For example polymers of $100$ monomers starting from initial conformations, whose monomer density is $0.001$, eject with the probability $96$ \%, whereas no ejection takes place through the symmetrical pore for this density, and ejections are recorded with the probability $0.06$ \% at $\rho = 0.2$. So, only a minor modification in the pore region is needed for viral ejection from a capsid to be initiated and completed.

The dynamics of the ejection through the asymmetrical pore resembles strongly that of forced translocation. Accordingly, we compare capsid ejection through the asymmetrical pore to polymer translocation through the symmetrical pore using identical Langevin dynamics and polymer models. The scaling reported for polymer translocation does not represent any supposedly ``correct'' value but is the scaling obtained for a translocation model that is as closely identical to the used capsid ejection model as possible. Indeed, for example due to finite-size effects, no universal scaling exponents can be directly obtained for polymer lengths amenable for simulations.

\begin{figure}[h]
\centerline{
\includegraphics[angle=0, width=0.213\textwidth]{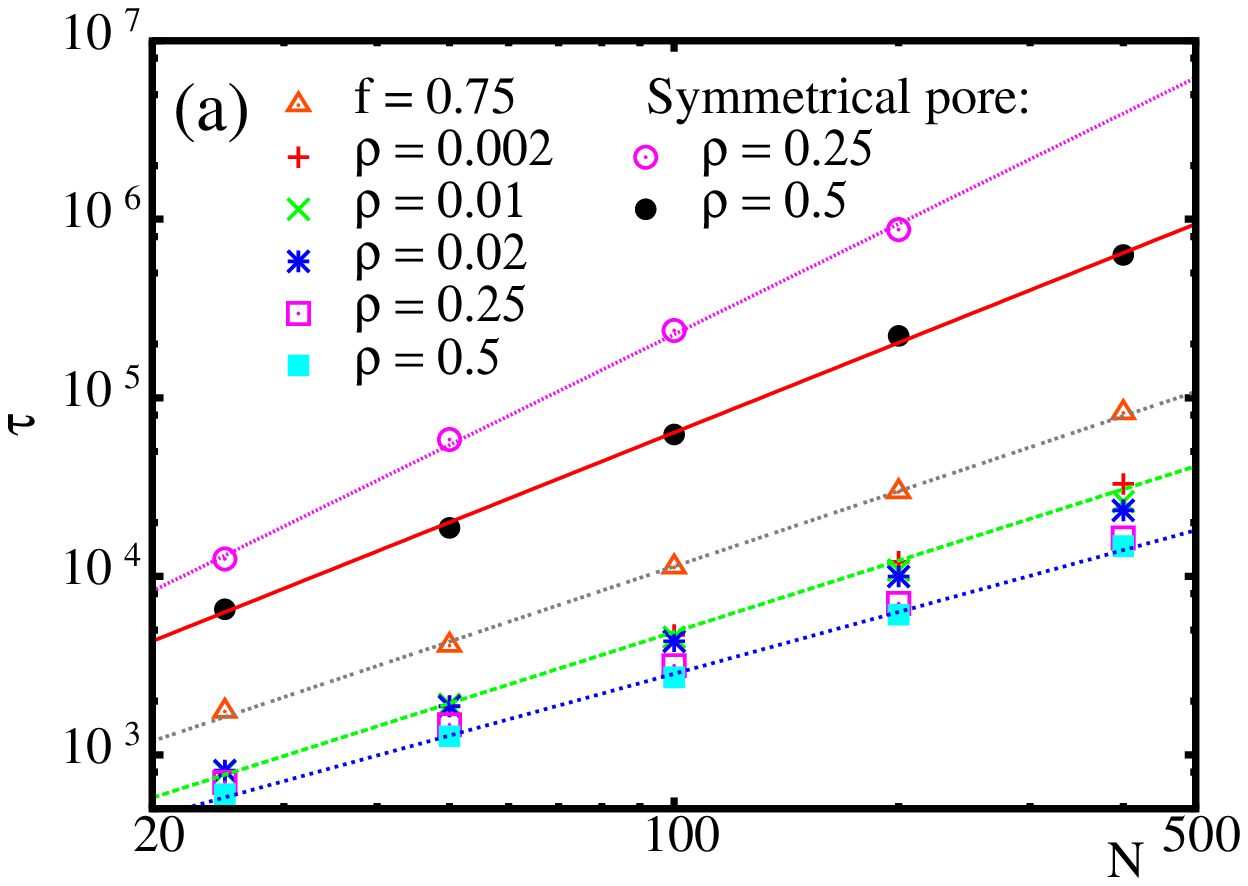}
\includegraphics[angle=0, width=0.213\textwidth]{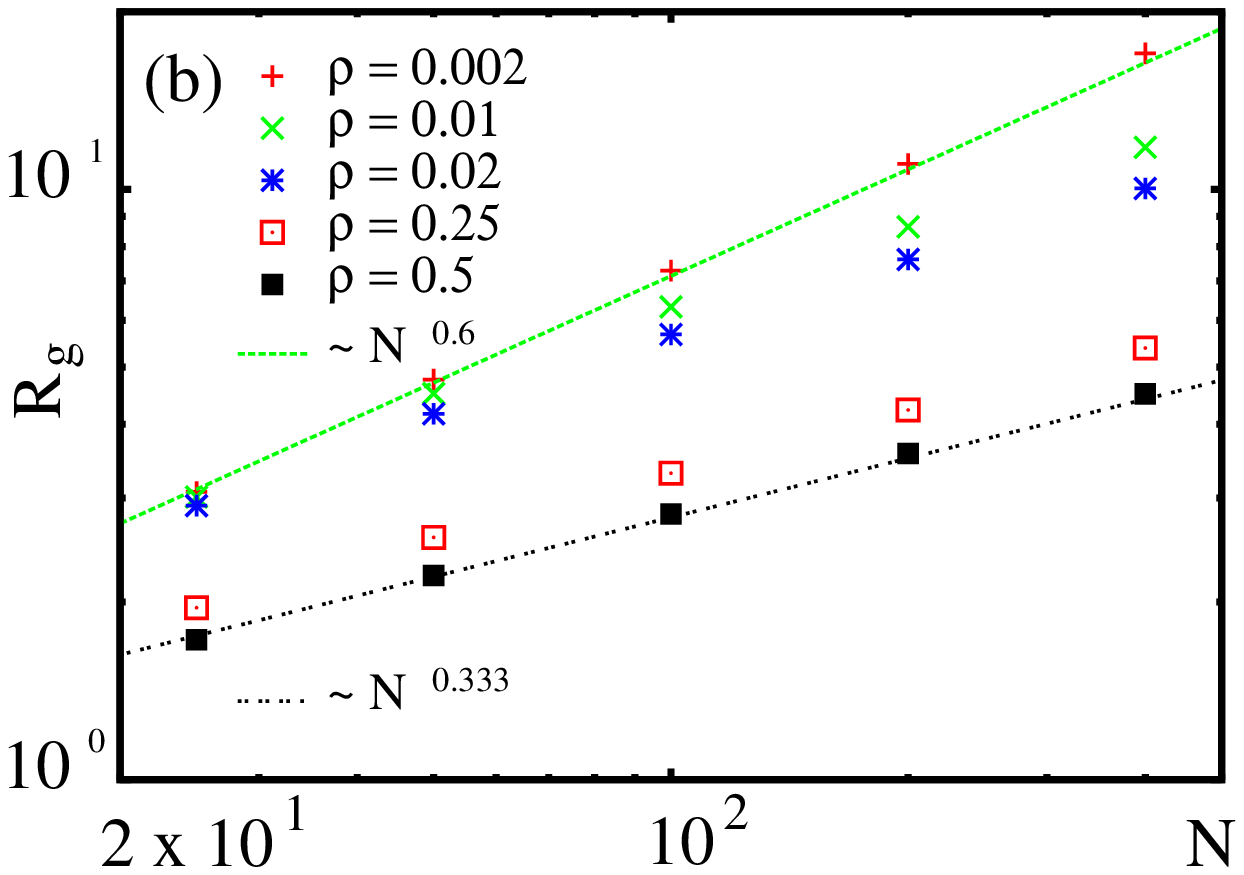}
}
\caption{(Color online) (a) Ejection and translocation times with least-squares fitted scaling relations $\tau \sim N^\alpha$. From the top down: Ejection times for capsids with the symmetrical pore, $\rho =0.25$ and $0.5$; $\alpha \approx 2.06$ and $1.67$, respectively. Polymer translocation driven by a pore force $f = 0.75$; $\alpha \approx 1.4$. Ejection times for capsids with the asymmetrical pore. $\alpha \approx 1.33$, $1.25$, and $1.21$ for $\rho = 0.002$, $0.01$, and $0.02$, respectively. Fitting would give $\alpha \approx 1.15$ and  $1.13$ for  $0.25$ and $0.5$, respectively.  However, for the asymmetrical pore the scaling breaks down for $\rho \gtrapprox 0.25$ (see the text). (b) Radii of gyration (in units $b = 1$) as a function of lengths $N$ for polymers initially inside the capsids. Initial densities are from top down as in (a). The scaling $R_g \sim N^\nu$ with $\nu = 0.6$ and $0.333$ are given for reference.}
\label{fig2}
\end{figure}

Ejection times are plotted for polymers of different lengths starting from different initial monomer densities $\rho$, see Fig.~\ref{fig2}(a). The ejection time $\tau$ is seen to scale with $N$ for all $\rho \lessapprox 0.25$. The translocation times for the case where polymers are driven through a pore in the wall dividing the space are given for reference. Here, the polymers were driven by applying a constant force $f$ on the segments inside the pore. For the pore force $0.25$ per bead, that is, the total pore force $f = 0.75$ the translocation time scales as $\tau \sim N^\alpha$, where $\alpha = 1.4$. The scaling turns out to be essentially the same for $f = 1.5$ and $3$ (not shown). This can be compared with capsid ejection, where for the lowest initial densities used, $\rho = 0.001$ and $0.002$, scaling is obtained with $\alpha \approx 1.33$. For these densities the radii of gyration for the polymers inside the capsids scale with the polymer lengths as $R_g \sim N^\nu$, where $\nu = 0.6$, see Fig.~\ref{fig2}(b). This is the scaling obtained for free polymers in a good solvent. Hence, the capsid confines the polymers only weakly at densities $\rho \le 0.002$, and the ejection time scales with the polymer length for capsid ejection almost as the translocation time for the polymer translocation using identical dynamics. 

The scaling form $\tau \sim N^\alpha$ with the smallest deviations was obtained for $\rho = 0.01$ and $0.02$ with $\alpha = 1.25$ and $1.21$, respectively. The dependence of $R_g$ on $N$ deviates from the free polymer scaling at these densities. For the asymmetrical pore the scaling breaks down at densities $\rho = 0.25$ and $0.5$. Power-law fitting would give the scaling exponent $1.1 < \alpha < 1.2$. Hence, in capsid ejection the scaling breaks down for $\rho \gtrapprox 0.25$. Moreover, for densities $\rho \gtrapprox 0.25$ the radius of gyration scales with the polymer length as $R_g \sim N^{0.333}$, so $\rho \approx 0.25$ marks the density beyond which the polymer is completely confined by the capsid. Apart from the different scaling exponent $\alpha$, the translocation time magnitudes for the driving force $f = 3$ (not shown) closely correspond to the ejection times for $\rho = 0.25$.

For reference, we implemented the pore asymmetry also in our translocation model. There, in the absence of the confinement due to the capsid, the asymmetry corresponds to the driving force $f \approx 0.83$ applied inside the symmetrical pore. The characteristics of these two cases were identical. Fitting gives values $\alpha = 1.36$ and $1.37$ for the unforced translocation with the asymmetrical pore and driven translocation, $f = 0.83$, with the symmetrical pore. For these translocations the ratio of polymers sliding back to the {\it cis} side are $5.5 \%$ and $21.3 \%$, respectively, so even though the characteristics for these two processes are identical, the asymmetry is more effective than pore force as a means of ensuring that the translocation is completed. 

Polymer translocation and polymer ejection from a capsid share the common feature of the translocation or the ejection time apparently scaling with polymer length. While the characteristics of the driven translocation seem to be largely determined by the spreading of tension on the {\it cis} side part of the polymer~\cite{sakaue,linna1,linna2,dubbeldam,ikonen,grosberg2}, a detailed conclusive explanation of how the scaling $\tau \sim N^\alpha$ changes with applied force is still missing. In capsid ejection, where the polymer is initially in a packed conformation, tension spreading inside the capsid can hardly explain the obtained scaling, at least for higher monomer densities. As pointed out in~\cite{sakaue3}, in capsid ejection the time-dependent radially transmitted pressure drop has a role analogous to tension spreading in polymer translocation.


\subsection{Event distributions}

In polymer translocation we have found that registering transitions in the process at small length scales reveals scaling in the numbers and times of such transitions and shows characteristics of a multiplicative stochastic process~\cite{linna3}. It is of interest to see if capsid ejection can be characterized similarly.  For this we observe the ejection process at a finer scale than the bond length $b$ and register polymer segment motion inside the pore with resolution $b/10$. In order to characterize the ejection process we define an event in this process as a transition of a polymer segment $\Delta s$ in either direction inside the pore without reversal. Hence, an event is not terminated by {\it e.g.}\ pausing the motion but only when the segment starts moving in the direction opposite to the previous direction. Defining an event in this way ensures that all the pauses will also be included just as they are in the definition of the total ejection time. The resulting distributions should not be confused with distributions of total ejection or translocation times reported {\it e.g.} in~\cite{milchev,cacciuto2}.

\begin{figure}[h]
\centerline{
\includegraphics[angle=0, width=0.215\textwidth]{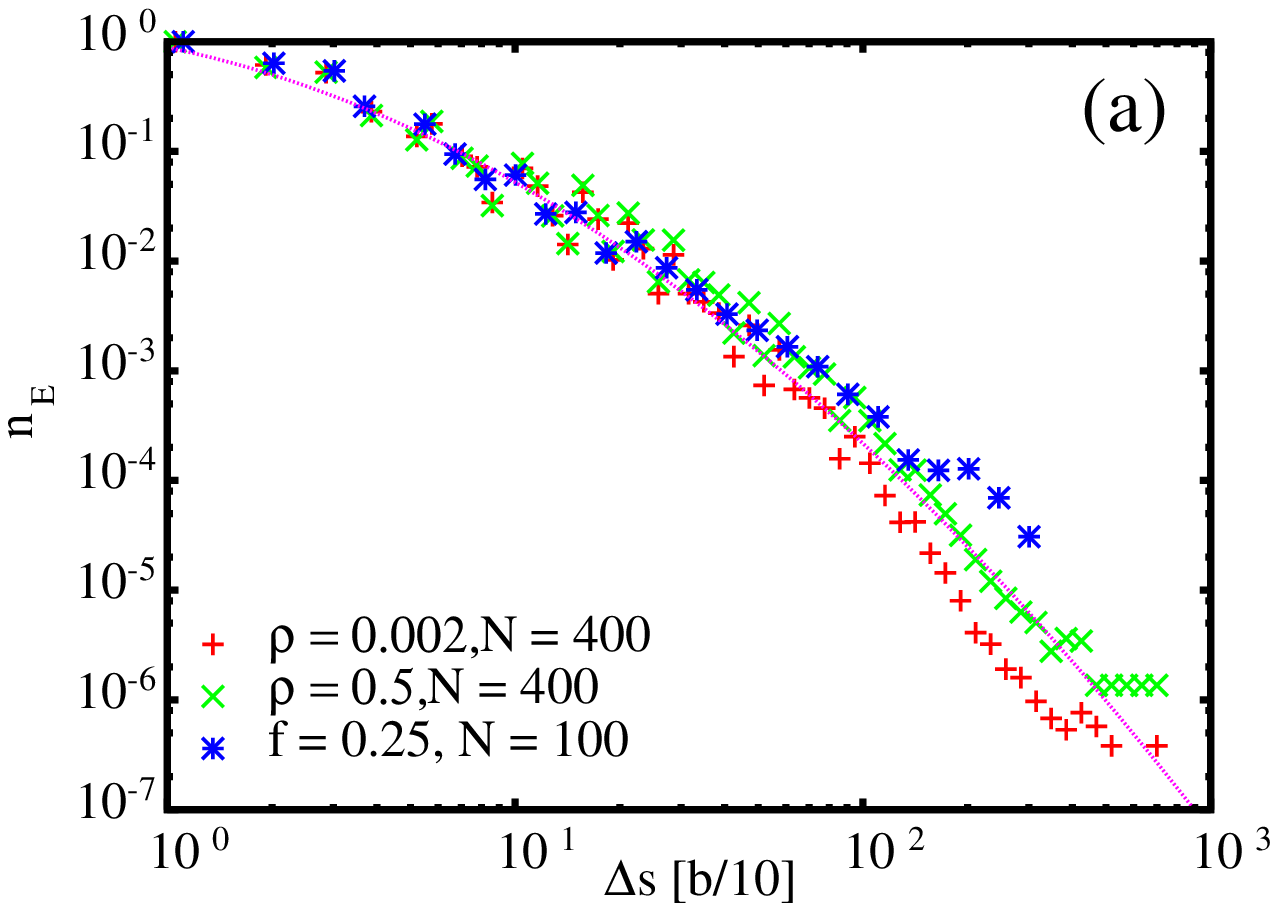}
\includegraphics[angle=0, width=0.215\textwidth]{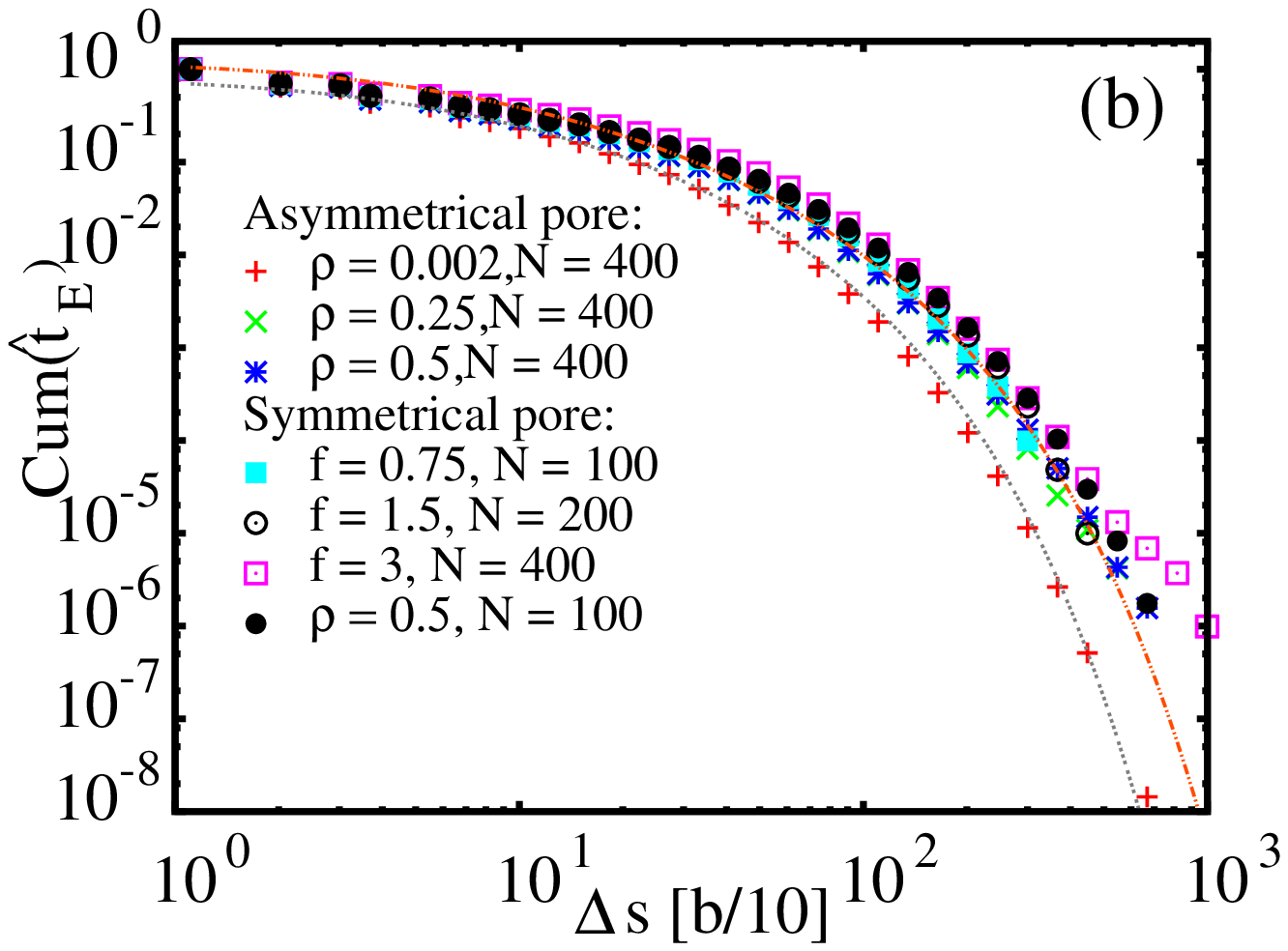}
}
\caption{(Color online) (a)~Asymmetrical pore. Numbers of events logarithmically binned on traversed segment lengths $\Delta s$ and normalized by the maximum numbers of events $\hat{n}_E$ for capsid ejection starting from monomer densities $\rho = 0.002$ and $0.5$ and driven translocation. The log-normal function $\hat{n}_E = (1/\Delta s) \exp\{-[\ln(\Delta s/2)]^2/4\}$ is plotted to guide the eye. (b)~Normalized cumulative event time distributions for capsid ejection through the asymmetrical pore with $N = 400$ and $\rho = 0.002$, $0.25$, and $0.5$ and through the symmetrical pore for translocation with $N = 100$, $f = 0.75$, and $N = 200$, $f = 1.5$ and for capsid ejection with $\rho = 0.5$ and $N = 100$. Functions to guide the eye are of the form given in Eq.~(\ref{acum}) with $\Delta s_0 = 6$. $\Delta s_c = 95$, and $\sigma = 1$  for the curve fitting cumulative event times for $\rho = 0.25$ and $0.5$, and $\sigma = 0.65$ for $\rho = 0.002$.}
\label{fig3}
\end{figure}

Figure~\ref{fig3} shows log-binned distributions of numbers of events $n_E$ for transitions of distance $\Delta s$. Ejection proceeds slowly in the case of the symmetrical pore and the distributions present large fluctuations exactly as was the case for unforced translocation~\cite{linna3}. Due to this, only cumulative distributions for the case of the symmetrical pore are shown in Fig.~\ref{fig3}(b). To compare their forms each noncumulative distribution is normalized by its maximum number of events, {\it i.e.}, $\hat{n}_E = n_E(\Delta s)/n_{max}$. The normalized distributions $\hat{n}_E(\Delta s)$ for capsid ejection are seen to be of log-normal form. This same form was obtained also for polymer translocation~\cite{linna3}. A distribution obtained for polymer translocation with $f = 0.75$ is shown in Fig.~\ref{fig3} for comparison.

Capsid ejection through both the symmetric and asymmetric pores can be characterized as a multiplicative stochastic process in the same way as polymer translocation. As the speed of ejection through the asymmetrical pore for $\rho = 0.25$ was found to be of the same order as the translocation speed for $f = 3$, it is natural to expect that stochasticity, that is fluctuations, play as important a part here as was found to be the case for driven polymer translocation~\cite{linna3}. The deterministic contribution in the polymer translocation comes from the force balance between the drag force and the  driving pore force. Tension spreading on the polymer contour on the $cis$ side determines the number of moving segments contributing to the drag force. As said, in capsid ejection tension spreading in the polymer confined inside the capsid is not likely to play an important role in the high density limit. Here the deterministic contribution comes from the higher pressure inside the capsid driving the ejecting polymer and, in the present case, by the bias coming from the pore asymmetry. In spite of these strong deterministic contributions in both processes, the dominance of the small-scale transitions implies that the effect of fluctuations cannot be neglected.

In both capsid ejection and polymer translocation multiplicative stochasticity arises naturally. We observe traversed polymer segments $\Delta s_n$, where $n$ indexes the registered segments. $\Delta s_n$ result from $m$ consecutive transitions of distance $s_i$. Denoting the probabilities of these individual transitions as $p_i$ ($1 \le i \le m$), the probability of $m$ such consecutive transitions can be written as $P^{(m)}_r = \prod_{i=1}^m p_i$. Now, $\ln P_r^{(m)} = \sum_{i=1}^m \ln p_i$. Due to the central limit theorem  $\ln P_r^{(m)}$ becomes a normal distribution with large $m$~\cite{redner}. Hence, $P^{(m)}_r$ becomes a log-normal distribution. Now, assuming that on average, {\it i.e.}\ averaging over different realizations of a sequence of $m$ events, $\langle \prod_{i=1}^m p_i \rangle \sim m$, the distribution of the number of events as a function of $m$ is of the same form as $P^{(m)}_r$, $P(m) \sim P^{(m)}_r$. Since $\Delta s_n = m s_i$ and, as will be seen, short-range transitions dominate the capsid ejection and translocation processes, $P(\Delta s_n) \sim P(m)$. A mathematically more rigorous justification for this last relation would be in order. This would require extensive data on $p_i$ and $s_i$ at different stages of capsid ejection and translocation.


The distributions of small-scale transitions obtained from the simulations are a good way to characterize the two processes. In what follows, we drop the subscript $n$ and use the symbol $\Delta s$ to denote the lengths of the traversed segments registered with resolution $b/10$. Log-normal distributions of the form

\begin{equation}
P(\Delta s) \sim \frac{1}{\Delta s} \exp \{- [\ln(\Delta s/\Delta s_0)]^2/2 \sigma^2\},
\end{equation}
were obtained for both capsid ejection and polymer translocation. Here,
$\Delta s_0$ is the characteristic scale and $\sigma$ the standard deviation of the variable $\ln \Delta s$, see {\it e.g.}~\cite{sornette}.

To relate the event distributions more directly to the obtained scaling $\tau \sim N^\alpha$ we measure the time $\Delta t(\Delta s)$ elapsed when transferring polymer segments of different lengths $\Delta s$ through the pore. We define $t_E$ as the total elapsed time 
in transitions of (constant) distances $\Delta s$ as 

\begin{equation}
t_E(\Delta s) = \sum_{\Delta t(\Delta s)} n_E(\Delta s, \Delta t) \Delta t(\Delta s),
\label{evtimes}
\end{equation}
where the sum runs over all measured $\Delta t$ for each constant $\Delta s$. For both capsid ejection and translocation we obtain distributions of the form

\begin{eqnarray}
P(t_E(\Delta s)) \sim \frac{1}{\Delta s} \exp \{- [\ln(\Delta s/\Delta s_0)]^2/2 \sigma^2\} \nonumber \\ 
\times \exp(- \Delta s/ \Delta s_c),
\end{eqnarray}
where $\Delta s_c$ is a finite cut-off. Due to the fluctuations present at small $\Delta s$ we show in Fig.~\ref{fig3}(b) cumulative distributions that are of the form 

\begin{eqnarray}
\label{acum}
P_C(t_E(\Delta s)) &=& \sum_{\Delta s = {+\infty}}^0 P(t_E(\Delta s)) \\ 
                   &\sim& 1 - \textrm{erf}\{ \sqrt{[\ln(\Delta s/\Delta s_0)]^2/2 \sigma^2 + \Delta s/ \Delta s_c} \} \nonumber .
\end{eqnarray}

Each cumulative distribution shown in Fig.~\ref{fig3}(b) is again normalized by its maximum value. The cumulative distributions for driven translocation with $f = 0.75$ and $1.5$ are given for comparison. These can be fitted using the form of Eq.~(\ref{acum}) with $\sigma = 7$, $\Delta s_0 = 6$. The data for capsid ejection can be roughly fitted with $\Delta s_0 = 6$ and $\Delta s_c = 95$.  For the asymmetrical pore $\sigma \approx 2$ for $\rho = 0.002$ and $\sigma < 1$ for $\rho = 0.25$ and $0.5$. So, analogously to increasing $f$ in translocation~\cite{linna3,try}, increasing $\rho$ in capsid ejection broadens the log-normal distributions, {\it i.e.} increases the dispersion $\sigma$.

Due to the bias induced by the pore asymmetry it is reasonable that the event distributions in the case of the asymmetrical pore should resemble those for the driven translocation. In~\cite{linna3} the forms of the cumulative distributions for unforced translocation were seen to closely resemble those for the driven translocation. Analogously, the cumulative distribution for the symmetrical pore resembles closely that for the asymmetrical pore.

\begin{figure}
\centerline{
\includegraphics[angle=0, width=0.215\textwidth]{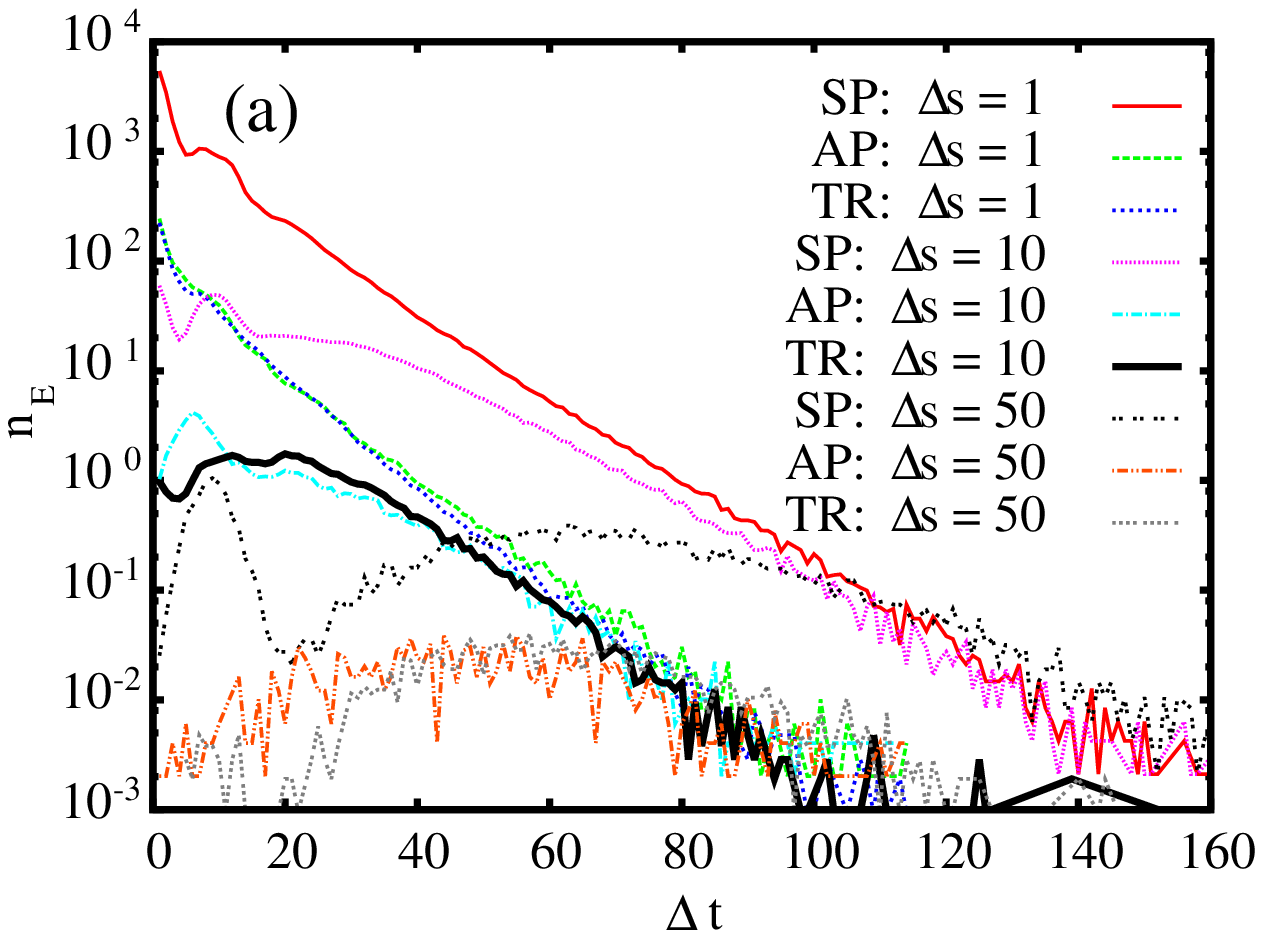}
\includegraphics[angle=0, width=0.215\textwidth]{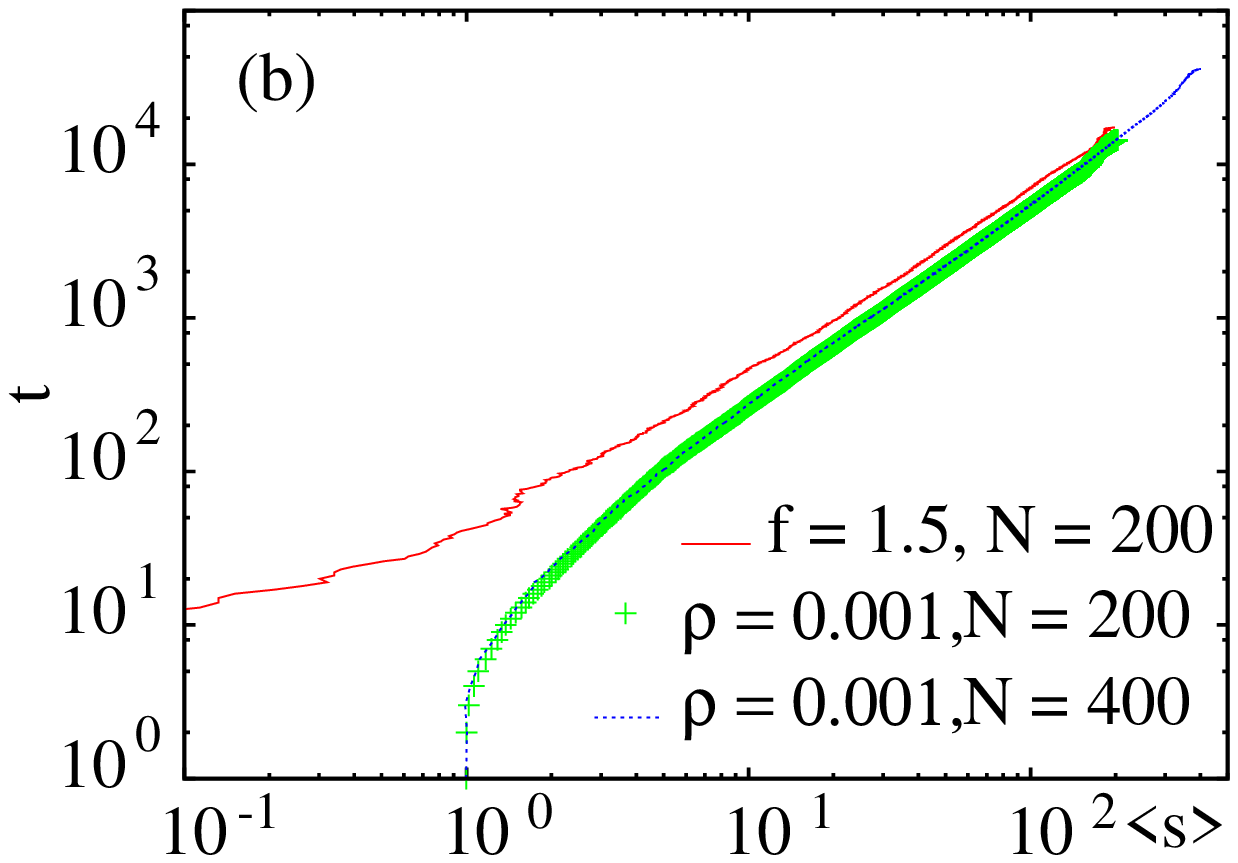}
}
\caption{(Color online) (a) The distributions of $n_E(\Delta t)$ for transitions of constant distances $\Delta s = 1$, $10$, and $50$ in capsid ejection with initial segment density $\rho = 0.5$ and in translocation with $f = 3$. For translocations (TR) the symmetrical pore was used. For capsid ejections both symmetrical (SP) and asymmetrical (AP) pores were used. $N = 100$. (b) The elapsed time plotted as a function of the reaction coordinate sampled at time intervals and averaged over individual simulations $\langle s \rangle$ for driven translocations for $f = 1.5$ and capsid ejections through the asymmetrical pore for $\rho = 0.001$ and $N = 200$ and $400$.}
\label{fig4}
\end{figure}

Extracting events of different segment lengths $\Delta s$ characterizes the roles of short and long range events in the capsid ejection. In Fig.~\ref{fig4}(a) we show the numbers of events $n_E$ taking time $\Delta t$ for transferring segments of lengths $\Delta s = 1$, $10$, and $50$ during capsid ejection and translocation. As in polymer translocation, short-ranged transitions are seen to dominate also in capsid ejection for both the asymmetrical and symmetrical pores. The distributions $n_E(\Delta t)$ for the capsid ejection are very much like those for the translocation.

For the asymmetrical pore and $\rho = 0.01$ the exponentially decaying $n_E(\Delta t)$ for $\Delta s = 1$ are shown in Fig.~\ref{fig5}(a). The maxima of $n_E(\Delta t)$ for $\Delta s = 1$ versus polymer length $N$ scale as $\textrm{max} (n_E) \sim N^{1.295}$; see Fig.~\ref{fig5}(b). The same scaling is obtained for $t_E(\Delta s)$ for constant $\Delta s$ (not shown), again in keeping with what was obtained for translocation~\cite{linna3}. This can be compared with the measured $\alpha = 1.25$ [Fig.~\ref{fig2}~(a)]. For the symmetrical pore the maxima $n_E(\Delta t)$ for $\Delta s = 1$ give the obtained scaling $\alpha \approx 1.7$.

Recapitulating, the event distributions show that capsid ejection dynamics through both the symmetrical and asymmetrical pore are dominated by short-scale transitions just as is the case for unforced and forced polymer translocations. In other words, fluctuations are important in both processes, which for example complicates the application of these processes in DNA sequencing. Based on the log-normal forms of the distributions the processes can be regarded as multiplicatively stochastic processes.

\subsection{Waiting time profiles}

We have previously investigated the waiting time profiles when analyzing the effect of tension spreading in the polymer segment on the {\it cis} side during translocation~\cite{linna2}. The waiting time profile gives the time each bead has to wait before it exits the pore. Integrating this over time should show at least some reminiscence of scaling in order for the relation $\tau \sim N^\alpha$ to result. First, we compare waiting times for ejection through the asymmetrical pore and polymer translocation. To obtain high resolution in $s$ we compute cumulative waiting time profiles from the simulations at constant time intervals by registering the value of the reaction coordinate at the pore exit, averaging the numbers obtained over individual runs, and plotting them as a function of time, as depicted in Fig.~\ref{fig4}~(b) for translocation and capsid ejection through the asymmetrical pore. The last translocated monomers were excluded due to different total translocation times causing large fluctuations to the average cumulative distribution obtained in this way.

\begin{figure}
\centerline{
\includegraphics[angle=0, width=0.215\textwidth]{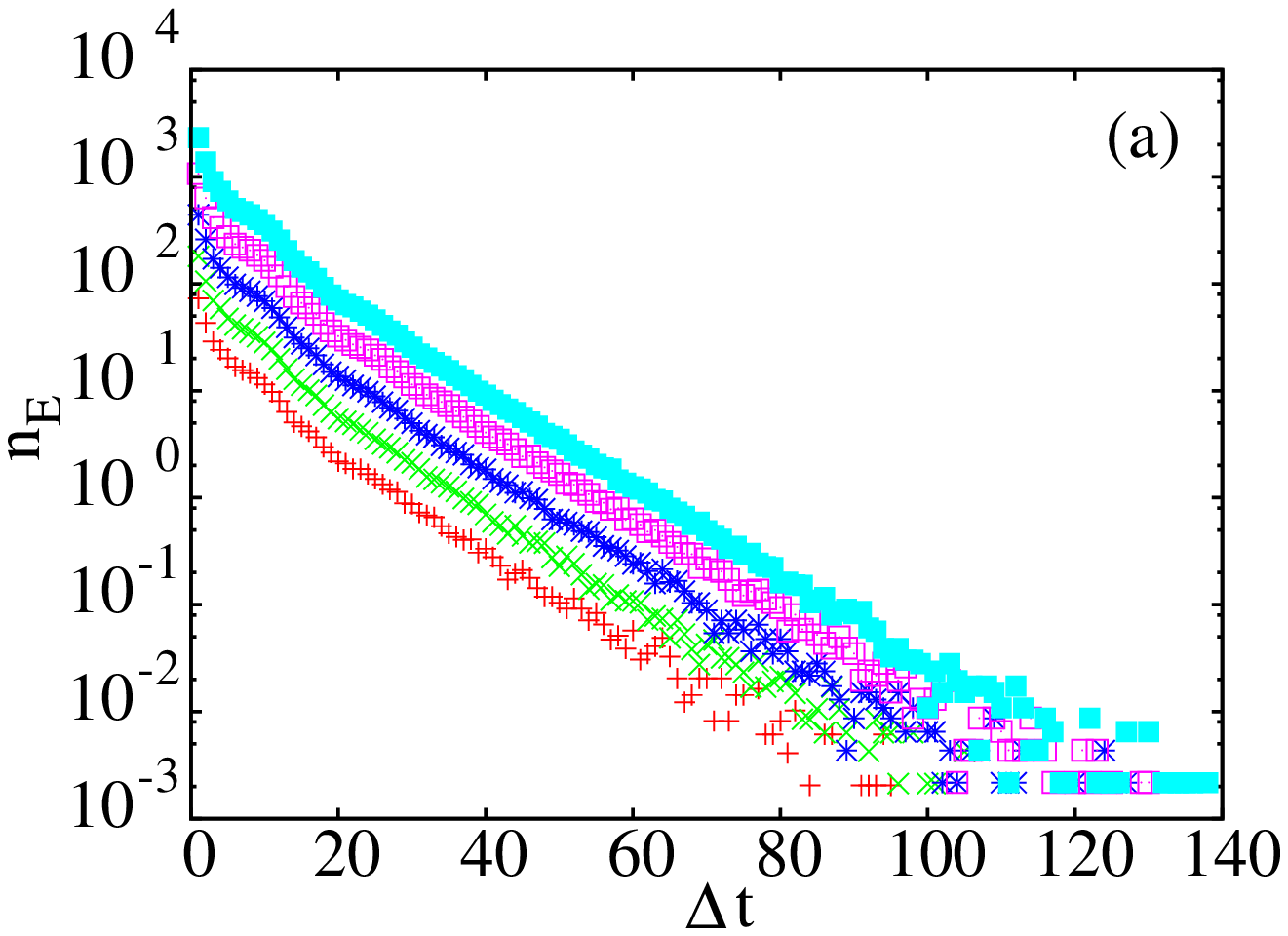}
\includegraphics[angle=0, width=0.215\textwidth]{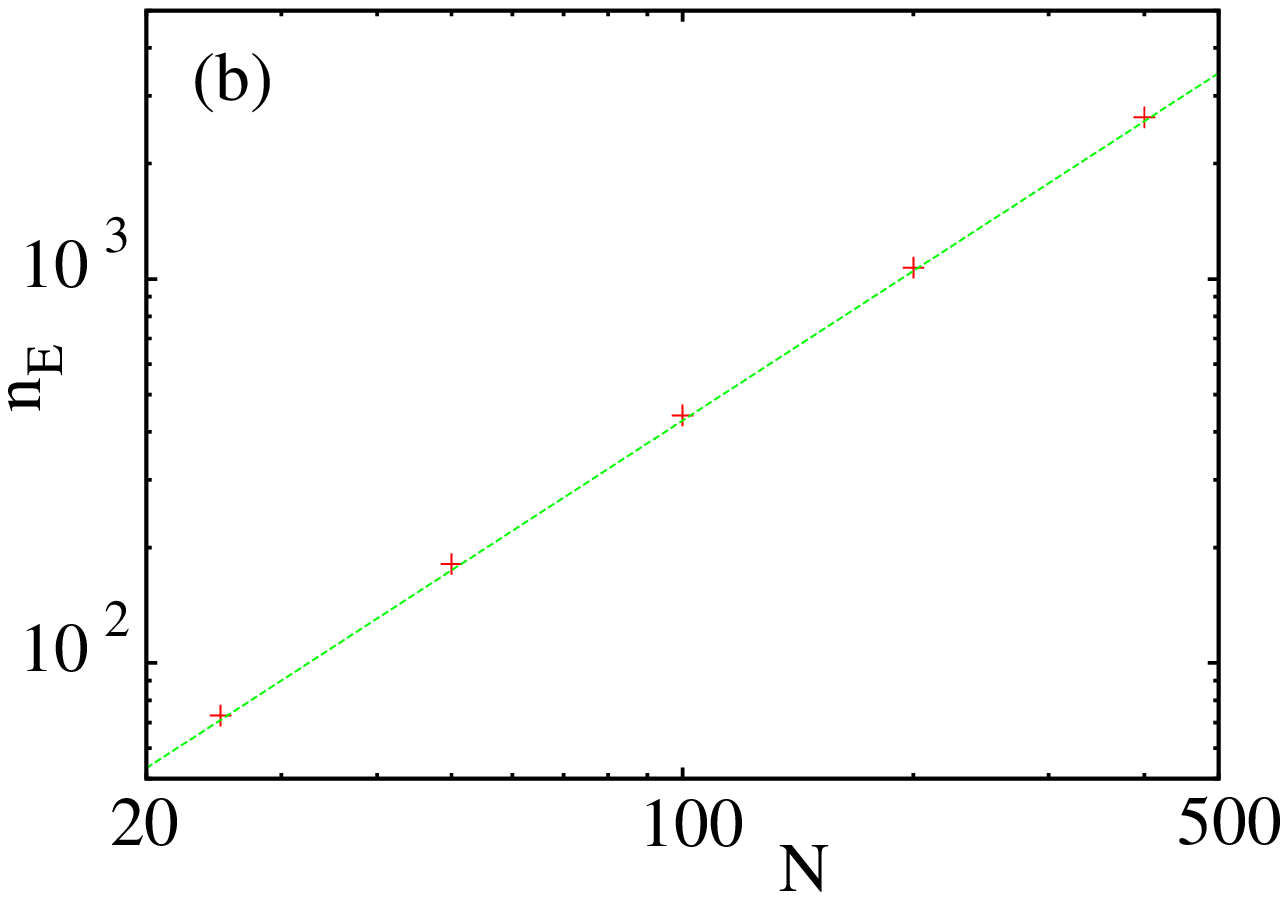}
}
\caption{(Color online) Asymmetrical pore. (a) The distributions of $n_E(\Delta t)$ for $\rho = 0.01$ and $\Delta s = 1$ for polymers of lengths $N = 25$, $50$, $100$, $200$, and $400$. (b) Maxima of the distributions $n_E(\Delta t)$ in (a) as a function of $N$. The function $n_E \sim N^{1.295}$ is plotted to guide the eye.}
\label{fig5}
\end{figure}

The resulting high-resolution cumulative waiting times for capsid ejection and driven translocation differ most in the beginning where capsid ejection is faster. This is because in capsid ejection the pressure inside the capsid pushing the polymer through the pore is greatest in the beginning, whereas in translocation the entropy difference opposes the process most effectively at the start. Towards the end the cumulative profiles become increasingly aligned. Indeed, the final stages, where $\rho$ is low, are very similar for the two processes. For the capsid ejection the cumulative waiting times scale with the sampled reaction coordinates as $t \sim s^\gamma$, where $\gamma \approx 1.2$. This is roughly the scaling found also for the number of events $n_E$ vs $N$, Figs.~\ref{fig5}~(a) and (b).  

\begin{figure}
\centerline{
\includegraphics[angle=0, width=0.215\textwidth]{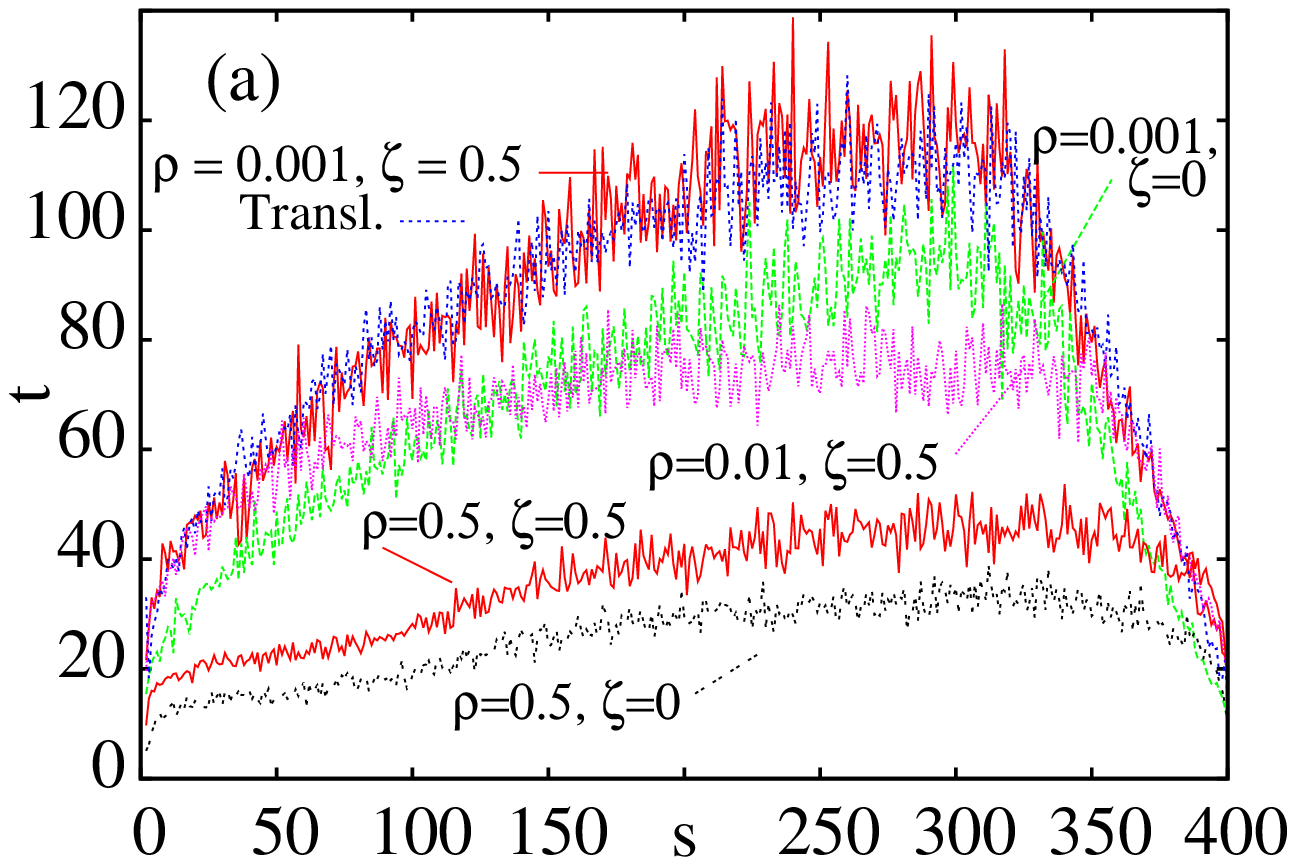}
\includegraphics[angle=0, width=0.215\textwidth]{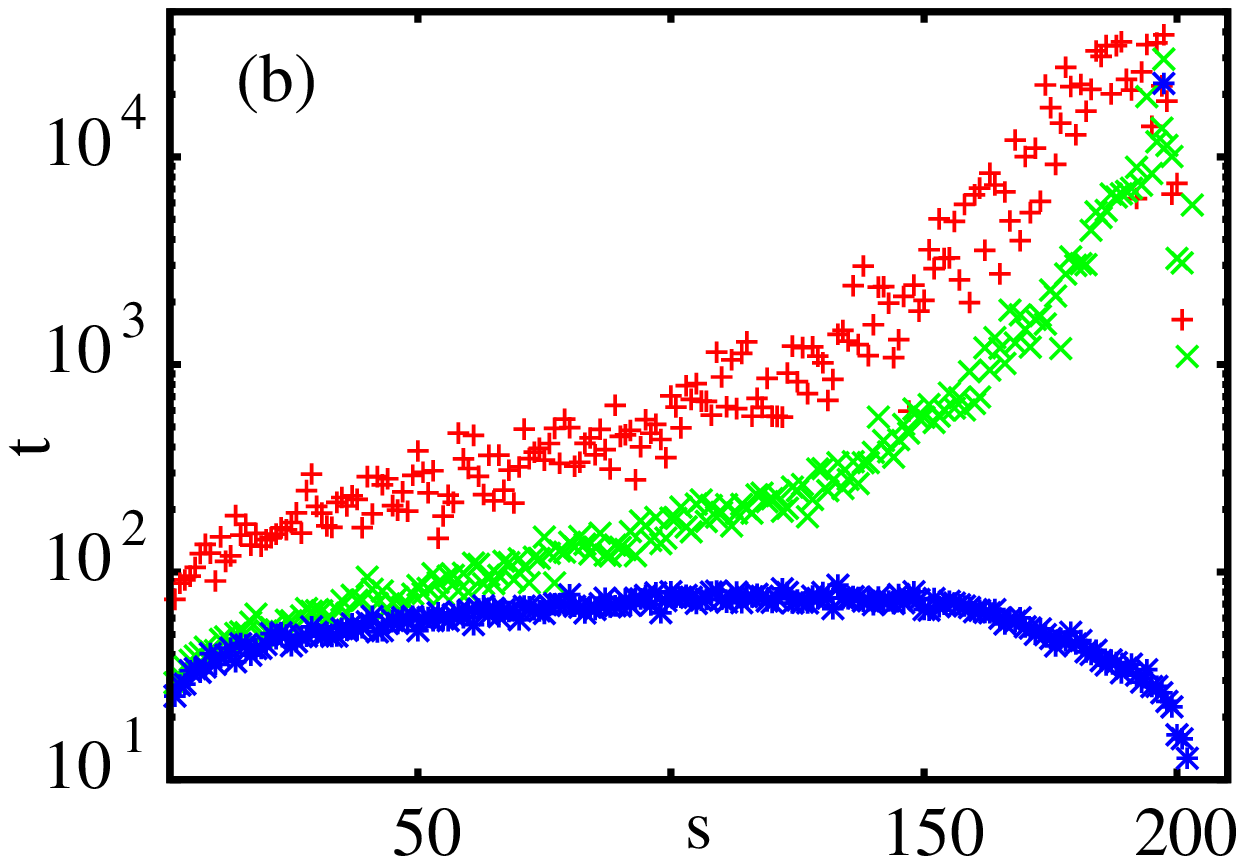}
}
\caption{(Color online) (a) Waiting time profiles for capsid ejection through the asymmetrical pore. Scaled (see text) driven polymer translocation is given for reference. Polymer length $N = 400$. The two topmost aligned curves: Driven translocation through the symmetrical pore for $f = 1$ (blue) and capsid ejection for $\rho = 0.001$ and $\zeta = 0.5$ (red). Below these curves from top down: capsid ejections for $\rho = 0.001$ and $\zeta = 0$ (green), $\rho = 0.01$ and $\zeta = 0.5$, and $\rho = 0.01$ and $\zeta = 0$. (b) Waiting time profiles from top down: Symmetrical pore $\rho = 0.25$ and $\rho = 0.5$, asymmetrical pore $\rho = 0.001$. Polymer length $N = 200$.}
\label{fig6}
\end{figure}

At a coarser scale, where the minimum registered displacement is the bond length, the waiting time forms of capsid ejection through the asymmetrical pore for $\rho = 0.001$ and driven translocation appear similar, see Fig.~\ref{fig6}~(a). The waiting times for the translocation that are of the same order as for the ejection starting from this low monomer density have been scaled so that the forms of the waiting time profiles can be compared. The waiting times for the driven translocation and the capsid ejection are in fact much more similar than was found for a computational tension spreading model and simulated driven polymer translocation (Fig.~2 in~\cite{ikonen}). Still, the values of $\alpha$ for the ejection and translocation processes here clearly differ. It can be concluded that very small differences in the forms of the waiting time profiles give rise to perceptible differences in the obtained scaling $\tau \sim N^\alpha$. Also, a conclusion of detailed similarity of processes or models cannot be made based on approximate similarity of waiting time profiles.

For increasing monomer density the waiting time profile is seen to flatten, which corresponds to a smaller scaling exponent $\alpha$. The rising slope associated with the tension spreading diminishes. In polymer translocation the finite-size effects can be viewed as the contribution from pore friction that is sufficiently strong compared with the overall friction. We will report our findings on the polymer translocation elsewhere~\cite{suhonen}. In capsid ejection the waiting time profiles do not change appreciably when pore friction $\zeta$ vanishes. Only the ejection times are reduced [see Fig.~\ref{fig6}(b)], but the scaling $\tau \sim N^\alpha$ (not shown) does not change when the pore friction is set close to zero.


Figure~\ref{fig6}~(b) shows waiting time profiles for ejections through the symmetrical and the asymmetrical pore. At the start the profile for $\rho = 0.25$ and the symmetrical pore and for $\rho = 0.001$ and the asymmetrical pore are closely aligned. Hence, the small pore asymmetry enhances the initial ejection of the polymer roughly as much as increasing the initial monomer density by two orders of magnitude. As $\rho$ diminishes from its initial value the ejection through the symmetrical pore slows down dramatically, whereas the asymmetry in the pore keeps the ejecting polymer in motion. This finding is in accord with the results for binding particles in~\cite{zandi}, where it was noted that the binding process involves important nonequilibrium effects resulting in the polymer being driven by a resultant force instead of just diffusing between binding sites in the Brownian ratchet fashion. 

\subsection{Scaling for capsid ejection through the asymmetrical pore}

Given that the pore asymmetry induces a small bias, we can sketch a crude derivation for the scaling obtained. If we ignore the pressure drop that is transmitted radially from the pore during the ejection and the tension that propagates along the chain at the final stages of the ejection, we may relate the total energy available for ejecting the polymer and the total energy dissipated in friction analogously to the estimate in~\cite{vocks} for driven translocation. In this way we get a lower limit estimate for the scaling $\tau \sim N^\alpha$.


The energy $E_{drive}$ driving the polymer out of confinement decreases during the ejection. It is initially $E_0$. The work done during ejection is $W = E N \le W_{max} = E_0 N$. For the densities $\rho = 0.01$ and $0.02$ the capsid confines the polymer, Fig.~\ref{fig2}(b). Accordingly, we take the number of moving monomers as roughly constant. This approximation is supported by the flattened waiting time profiles for $\rho = 0.01$, see Fig.~\ref{fig6}~(a). Since the capsid fully confines the polymer only for $\rho \gtrapprox 0.25$ [see Fig.~\ref{fig2}(b)], we take the distance traveled by a monomer during ejection to be of the order $\sim R_g$ for $\rho$ in the order of $0.01$. Then the average monomer velocity is given by $v_m \sim R_g/\tau$. Hence, the energy dissipated in friction during the whole ejection can be written as $E_{drag} \sim N \tau \xi v_m^2 = N \xi \frac{R_g^2}{\tau}$. From the relation $E_{drive} = E_{drag}$ the lower limit for the ejection time for  $\rho \approx 0.01$ can be written as

\begin{equation}
\tau \sim R_g^2/E > R_g^2/E_0 \sim N^{2 \nu} = N^{1.2}.
\end{equation}

The scaling exponent $\alpha = 2 \nu = 1.2$ for this crude lower-limit estimate is close to what was obtained for $\rho = 0.01$ and $0.02$. The estimate is based on the assumption that due to initial confinement the polymer ejects from the spherical conformation without forming a trumpet consisting of blobs~\cite{grosberg2}. Hence, the (average) length of the moving segment and so the drag experienced by the polymer are taken as constants. This lower limit fits the measured scaling exponents obtained for monomer densities that are smaller than $\rho \approx 0.25$, beyond which scaling breaks down for ejection through the asymmetrical pore.

\section{Conclusion}
\label{concl}

We have studied the process of a polymer, initially in a random conformation, escaping from a spherical capsid through a nanoscale pore. We used two pore models, one of which has a minor asymmetry, namely, a small aligning force exerted on the monomer at the pore entrance inside the capsid. Although no bias is explicitly applied in the direction of ejection, the alignment imparts a preference to the direction of the ejection. Dynamically the situation is largely analogous to that of particles binding on the polymer and partly preventing it from sliding back~\cite{zandi}. In the case of capsid ejection such an asymmetry of the pore could result from the different curvatures of the protein capsid on the outer and inner capsid surfaces. This asymmetry is a very efficient way of initiating and completing the polymer ejection from a capsid. It is thus one candidate for such a mechanism.

The ejection process through the symmetrical pore was found to be very slow and very unlikely for initial monomer densities $\rho \lessapprox 0.25$. This is in stark contrast with the results of Monte Carlo simulations in~\cite{muthukumar1,cacciuto2}. Also, the predictions for the scaling $\tau \sim N^\alpha$ in these references were seen not to hold for our simulations. This we ascribe to nonequilibrium effects, which are well described in~\cite{sakaue3}. Qualitatively, capsid ejection results from a drift component due to high monomer density inside the capsid and diffusion. The scaling exponent $\alpha$ decreases with increasing $\rho$ for both the symmetrical and asymmetrical pores, which is in keeping with the drift dominating for large $\rho$.

The generic similarity of polymer translocation and capsid ejection was shown by investigating the distributions of short transitions. We showed that capsid ejection, analogously to polymer translocation~\cite{linna3}, shows characteristics of multiplicative stochastic processes. Capsid ejection showed very similar characteristics to polymer translocation. Stochastic multiplicity typically describes non-equilibrium processes. The dominance of short-ranged transitions in these processes shows that fluctuations play an important role in them, which has consequences for example for DNA sequencing. (For recent analyses on stochasticity and fluctuations in driven translocation, see~\cite{sakaue2} and~\cite{dubbeldam2}.)

The ejection through the asymmetrical pore was found to dynamically resemble driven polymer translocation, although the initial conformations in the two processes differ substantially. Both event distributions and waiting time profiles for these processes had very similar forms despite the fact that their scaling exponents differed. In the literature, fairly detailed conclusions are made based on waiting time profiles. These are not warranted by our findings. The correct scaling is, however, very precisely seen in the number of small-scale transition events. This can be viewed rather as a consequence of than an explanation for the scaling. 

We showed via measured radii of gyrations that the breakdown of the scaling relation occurs at monomer densities, for which the initial conformations are completely confined. In a similar way as was done for driven translocation in~\cite{vocks} we gave a crude estimate of the lower limit for the scaling of the ejection time with polymer length based on the energy available for ejection and the energy dissipated during ejection. This lower limit estimate of $\alpha = 1.2$ is in keeping with the results from our simulations. 

In conclusion, we have characterized the dynamics of the ejection and translocation processes and found that a small asymmetrically applied aligning force at the pore entrance provides a strong means of driving the polymer through a nanoscale pore. It may facilitate many biological polymer translocation processes. It can also find application in driving polymers through fabricated pores for analytical purposes {\it e.g.} in lab-on-a-chip applications.


\begin{acknowledgments}
The computational resources of Aalto Science-IT Project and CSC - IT Centre for Science, Finland, are 
acknowledged.
\end{acknowledgments}

\bibliography{capsid}

\begin{thebibliography}{47}
\bibitem [{try()}]{try}%
  \BibitemOpen
  \href@noop {} {}\bibinfo {note} {In~\cite{linna3} the dispersions given in
  Fig. 3 are in fact $2 \sigma^2$ instead of $\sigma$.}\BibitemShut {Stop}%
\end{thebibliography}

\end{document}